\documentclass[10pt,prd,aps,twocolumn,preprintnumbers,showpacs, nofootinbib,superscriptaddress,notitlepage]{revtex4-2}

\def \Dslash {D\!\!\!\!/}

\usepackage{slashed}
\usepackage{graphicx,color}
\usepackage{epsfig}
\usepackage{subfigure}
\usepackage{epsfig}
\usepackage{multirow}
\usepackage{epstopdf}
\usepackage{dcolumn}
\usepackage{bm}
\usepackage{color}
\usepackage{ulem}
\usepackage{enumitem}
\usepackage[colorlinks,
linkcolor=black,
anchorcolor=black,
citecolor=black
]{hyperref}

\begin{document}

	\baselineskip=15pt
	
	\preprint{CTPU-PTC-25-10}

\title{Lepton collider imprints of Inelastic dark matter  model}

\author{Wei Liu}
\email{wei.liu@njust.edu.cn}
\affiliation{ Department of Applied Physics and MIIT Key Laboratory of Semiconductor Microstructure and Quantum Sensing,
Nanjing University of Science and Technology, Nanjing 210094, China}

\author{Jin Sun}
\email{sunjin0810@ibs.re.kr(Contact author)}
\affiliation{Particle Theory and Cosmology Group, Center for Theoretical Physics of the Universe, Institute for Basic Science (IBS), Daejeon 34126, Korea }

\begin{abstract}
 We explore  the potential for detecting  the inelastic dark model (DM) with an additional  $U(1)_D$ gauge symmetry at various types of lepton colliders.
 The new gauge boson $Z'$ resulting from the spontaneous breaking of $U(1)_D$ 
 can act  as a portal connecting  the Standard Model fermions and DM fermions $\chi$, 
 thereby facilitating the production of $Z'$ and its decay into DM fermions.
The mass difference between the excited DM fermion $\chi_2$ and the ground DM fermion $\chi_1$ induces the subsequent decays $\chi_2\to \chi_1 +\text{vis}$, where $\chi_2$ and $\chi_1$ serve as  long lived particles and DM candidate, respectively.
For the inelastic DM with a mass $m_{\chi_1}$ in the range of (1-100) GeV, we find that the future lepton colliders can probe many regions not accessible by current experiments and provide greater projected sensitivity than the LHC and Belle in previous studies.
This suggests that future colliders can provide complementary exploration of the inelastic dark matter model.

\end{abstract}

\maketitle

\section{Introduction}

Although direct detection remains elusive, cosmological observations and early universe simulations provide indirect evidence for the presence of dark matter (DM), which is believed to constitute approximately 85\% of the total matter~\cite{Planck:2018vyg,ParticleDataGroup:2024cfk}.
Nonetheless, fundamental questions remain unresolved, such as the microscopic nature of dark matter (DM) and the origin of its observed relic density.
Numerous dark-sector models  have been proposed to explain the nature of the DM, including gauge bosons, scalar bosons, and fermions, see e.g.~\cite{Bernal:2017kxu,Ko:2018qxz,Argyropoulos:2021sav} for review.
In this framework, it is hypothesized that there exists a new sector of particles below the electroweak scale that are neutral under the Standard Model (SM) but could exhibit observable phenomena.

DM detection and identification at the microscopic level primarily involve direct detection (DD), indirect detection (ID), and collider experiments.
The first two types of searches rely on observing recoils from the elastic scattering of DM particles on the nuclei in the detector material or detecting annihilation products of DM pairs, such as monochromatic photons.
Collider searches for DM, on the other hand, rely on the production of DM particles in high-energy particle collisions and can be divided into two broad categories based on the experimental signatures they produce: (i) searches for missing transverse momentum plus X signatures for triggering, and (ii) searches involving only visible particles, such as pairs of leptons or jets, aimed at detecting the particles mediating interactions between DM and SM particles through the observation of new resonances or modifications in the kinematics of the final-state particles.

Among various DM models,  the inelastic DM model (iDM) has attracted significant attention, originally motivated by the explanation of the DAMA excess~\cite{Tucker-Smith:2001myb}.
A straightforward realization of the iDM scenario is that the DM particle, charged under a hidden gauge or global symmetry, experiences spontaneous symmetry breaking, resulting in symmetry-breaking masses much smaller than the symmetry-preserving ones, thus generating off-diagonal DM interactions.
Therefore, the particle components of the iDM scenario include the dark photon $Z'$ from the global $U(1)_D$ symmetry, a scalar $\Phi$ responsible for U(1) breaking to give the mass $m_{Z'}$, and new dark fermions $\chi_1$ and $\chi_2$, with the ground state $\chi_1$ serving as the DM candidate and $\chi_2$ as a long-lived particle.
Given the relatively long lifetime of $\chi_2$, 
a common method called the displaced vertex technique is used to detect long-lived particles via signatures~\cite{Antusch:2016vyf,BhupalDev:2018tox,Cheung:2019qdr,Schafer:2022shi,Urquia-Calderon:2023dkf,Lu:2024fxs}.
Further extensive studies and search strategies for the iDM model can be find in Refs.~\cite{Izaguirre:2015zva,Izaguirre:2017bqb,Berlin:2018jbm,Mohlabeng:2019vrz,Tsai:2019buq,Ko:2019wxq,Duerr:2019dmv,Duerr:2020muu,Kang:2021oes,Batell:2021ooj,Guo:2021vpb,Li:2021rzt,Filimonova:2022pkj,Bertuzzo:2022ozu,Gu:2022vgb,Mongillo:2023hbs,Mongillo:2023hbs,Heeba:2023bik,Lu:2023cet}, achieving sensitivity to the kinetic mixing $\epsilon \sim 10^{-3}$ around the inelastic DM mass range (1-100) GeV.
Here, $\epsilon$ denotes the kinetic mixing strength between the new $U(1)_D$ gauge field $X$ and the $U(1)_Y$ hypercharge field $B$, represented as $\epsilon/2 X_{\mu\nu} B^{\mu\nu}$.
 Note that the above investigations are primarily based on the hadron collider LHC and Belle.
Given the advantages of lepton colliders (clean environment, fixed collider energy, higher luminosity, etc.), they may provide much greater sensitivity to $\epsilon$, warranting further detailed exploration.
To the best of our knowledge, no relevant research has been conducted in this particular direction to date,
 except for a similar study involving exotic Z boson decays in the context of magnetic inelastic dark matter~\cite{Liu:2017zdh}. 
Since $\chi_1$ serves as the DM candidate required by the observed DM relic density, we focus on the mass range $m_{\chi_1}\in (1,\;100)$ GeV~\cite{Izaguirre:2015zva,Izaguirre:2017bqb,Duerr:2019dmv}.

In this paper, we investigate the potential of lepton colliders to probe the inelastic dark matter (iDM) model with an additional $U(1)_D$ gauge symmetry.
The new gauge boson $Z'$ acts as a portal connecting the visible electrons and the dark states $\chi_{1,2}$, facilitating the production of DM fermions~\footnote{The possibility of $Z'$ decays into right-handed neutrinos has also raised a lot of attention~\cite{Amrith:2018yfb,Deppisch:2019kvs,Deppisch:2019ldi,Liu:2021akf,Asai:2022zxw, Liu:2022kid,Li:2023dbs,Liu:2023klu,KA:2023dyz, Deppisch:2023sga,Liu:2024fey}.}.
Moreover, the mass splitting between the two dark states ($\Delta_\chi=m_{\chi_2}-m_{\chi_1}$), induced by the interaction between the dark Higgs field and the DM sector, leads to the decay of the excited state $\chi_2$ into the ground state $\chi_1$.

The rest of the paper is organized as follows. 
In section \ref{sec2}, we discuss  the details of  the  iDM model with  an $U(1)_D$ gauge symmetry. 
Section \ref{sec3} presents the collider signatures at different $e^+e^-$ colliders.  
In section \ref{sec4}, we present our results on the sensitivity to the iDM model parameters from different lepton colliders.
We provide our conclusions in section \ref{sec5}.

\section{Inelastic dark matter  model}\label{sec2}

In this section, we briefly review the inelastic dark matter  models with an $U(1)_D$ gauge 
symmetry,  focusing on fermionic DM candidates.
The  gauge group of   the model is $SU(3)_C\times SU(2)_L\times U(1)_Y\times U(1)_D$.
In addition to the SM particles, new particles are introduced: a singlet
complex scalar field $\Phi$ and  a Dirac fermion field $\chi$, as shown in Table.~\ref{tab:particle}.
All SM particles are
neutral under the $U(1)_D$ symmetry and cannot couple directly to the dark sector.
Note that  the $U(1)_D$ gauge symmetry is broken
into its $Z_2$ subgroup, with $Z_2-odd$ DM candidate  $\chi$, to ensure the DM stability.
The relevant particles and their charges are shown in Table.~\ref{tab:particle}

\begin{table}[htbp!]
\caption{The model particles and corresponding charges.
}
\begin{tabular}{c|c|c|c}\hline\hline
  &SM particles  &singlet scalar $\Phi$ & Dirac fermion $\chi$ \\\hline
$U(1)_D$ & 0 & +2 & +1 \\
$Z_2$ & +1 & +1 & -1 \\\hline
\end{tabular}
\label{tab:particle}
\end{table}

 The
relevant gauge invariant and renormalizable Lagrangian can be written as~\cite{Lu:2023cet}
\begin{eqnarray}\label{int}
\mathcal{L}&=&\mathcal{L}_{SM}-\frac{1}{4}X_{\mu\nu}X^{\mu\nu}-\frac{1}{2}\epsilon X_{\mu\nu}B^{\mu\nu}+\mathcal{D}^{\mu}\Phi^{\dagger}\mathcal{D}_{\mu}\Phi\nonumber\\
&-&\mu_\Phi^2\Phi^\dagger\Phi-\lambda_\Phi(\Phi^\dagger\Phi)^2-\lambda_{\mathcal{H}\Phi}\mathcal{H}^\dagger\mathcal{H}\Phi^\dagger\Phi\nonumber\\
&+&\overline{\chi}(i\mathcal{\Dslash}-M_{\chi})\chi-\left(\frac{\xi}{2}\Phi^{\dagger}\overline{\chi^{c}}\chi+H.c.\right)\;. 
\end{eqnarray}
Here the first line in Eq.~(\ref{int}) refers to the kinetic energy terms involving the $X_{\mu\nu}(B_{\mu\nu})$ field strength tensors of $U(1)_D(U(1)_Y)$  gauge fields.
$\epsilon$ means kinetic mixing, quantifying the Abelian mixing strength between $X_{\mu\nu}$ and $B_{\mu\nu}$.

The second line in Eq.~(\ref{int}) describes the scalar potential composed by complex doublet $H$ and singlet $\Phi$ as
\begin{eqnarray}
H&=& \left( \begin{array}{c}
      h^+\\\\ \frac{1}{\sqrt{2}}(v+h+i A)\end{array} \right), \nonumber\\
      \Phi&=& \frac{1}{\sqrt{2}}(v_X+h_X+iA_X)\;,
\end{eqnarray}
where $v(v_X)$  are vacuum expectation values of $H(\Phi)$,  which are responsible for breaking the  electroweak ($U(1)_D$) symmetries, respectively.
And $v=246$GeV. 
After symmetry breaking, the two scalar fields will mix each other to generate the physical fields as
    \begin{eqnarray}
	&&\left (\begin{array}{c}
		h\\
		h_X
	\end{array}
	\right )
	=
	\left (\begin{array}{cc}
		c_\alpha\;\;\;&\;\;\; s_\alpha\\
		-s_\alpha\;\;\;&\;\;\;c_\alpha
	\end{array}
	\right )
	\left (\begin{array}{c}
		H_1\\
		 H_2
	\end{array}
	\right ),\;
    \sin2\alpha=\frac{\lambda_{H\Phi}vv_X}{m^2_{H_2}-m^2_{H_1}}
\end{eqnarray}
here  $c_\alpha = \cos\alpha$, $s_\alpha = \sin\alpha$. In the following analysis, we adopt the same notation. we identify $H_1$ as the observed SM-like Higgs with $m_{H_1}=125$GeV.

The third line in Eq.~(\ref{int}) describes the fermion interaction relevant to  new Dirac fermion $\chi$, which are further decomposed into two Majorana fermion fields  $\chi_{1,2}$ as
\begin{eqnarray}
\chi=\frac{1}{\sqrt{2}}(\chi_2+i\chi_1),\quad \chi_2=\chi_2^c,\quad
\chi_1=\chi_1^c.  
\end{eqnarray}
After the $U(1)_D$ symmetry breaking, the DM interaction can be expanded as
\begin{eqnarray}\label{chi}
\mathcal{L}_{\chi}&=&\frac{1}{2}\sum_{n=1,2}\overline{\chi_n}(i \slashed{\partial}-M_\chi)\chi_n-i\frac{g_D}{2}(\overline{\chi_2}\slashed{X}\chi_1-\overline{\chi_1}\slashed{X}\chi_2)\nonumber\\
&-&\frac{\xi}{2\sqrt{2}}(v_X+h_X)(\overline{\chi_2}\chi_2-\overline{\chi_1}\chi_1)\;.
\end{eqnarray}
Furthermore, we obtain the $\chi_{1,2}$ mass as
\begin{eqnarray}
M_{\chi_{1,2}}=M_\chi\mp\frac{1}{\sqrt{2}}\xi v_X\;,
\end{eqnarray}
with the mass splitting parameterized as
\begin{eqnarray}
    \Delta_\chi=\sqrt{2}\xi v_X\;.
\end{eqnarray}

Generally,  the kinetic mixing $\epsilon X_{\mu\nu}B^{\mu\nu}$   in Eq.~(\ref{int}) can be removed by different ways: 
a)  $X_p$ in the canonical form does not
couple to hyper-charge current $j_Y^\mu$~\cite{Holdom:1985ag, Dobrescu:2004wz};
b)  the hyper-charge field in the canonical form $B_p$ does not couple to dark current $j^\mu_X$~\cite{Foot:1991kb, Babu:1997st};
c) in the most general scenario, an orthogonal transformation can be used such that, in the canonical basis, both the dark gauge boson $X_p$ and the hypercharge gauge field $B_p$ couple to $j_Y^\mu$ and $j^\mu_X$~\cite{Feldman:2007wj}.
Here we  choose case b) to conduct the analysis with  
the following transformation 
as~\cite{Sun:2023kfu}
\begin{eqnarray}\label{caseb}
	&&\left (\begin{array}{c}
	B\\\\
	X
\end{array}
\right )
=
\left (\begin{array}{cc}
	1 \;&\;\;\;-\eta\\\\
	0 \;&\;\;\; \frac{1}{\sqrt{1-\epsilon^2}}
\end{array}
\right )
\left (\begin{array}{c}
	B_p\\\\
	  X_p
\end{array}
\right )\;,
\end{eqnarray}
here we define $\eta=\epsilon/\sqrt{1-\epsilon^2}$.

The mass terms of gauge bosons are from the  scalar kinetic term given by
\begin{eqnarray}
	&&\mathcal{L}_{scalar} =  D^\mu \Phi^+ D_\mu \Phi + D^\mu H^+ D_\mu H, \;\;\;\mbox{with} \nonumber\\
	&&D_\mu =(\partial_\mu+ig' Y B_\mu+i g T_i  W_{i\mu} +ig_D Q_X X_{\mu})\;.
\end{eqnarray}
Here $g'$, $g$ and $g_D$ mean the $U(1)_Y$, $SU(2)_L$ and $U(1)_D$ gauge coupling constants, respectively.
Similarly, we need to adopt the transformation in Eq.~(\ref{caseb}) to express the fields in the canonical form $B_p$ and $X_p$.    
After electroweak symmetry breaking, we have 
\begin{eqnarray}\label{SM}
	B_{p\mu} = c_W \tilde A_\mu -s_W \tilde Z_\mu\;,\;\;  W^3_\mu =  s_W \tilde A_\mu +  c_W \tilde Z_\mu\;,
\end{eqnarray}
where  $\theta_W$ being the weak mixing angle. 
Moreover, the $\tilde Z$ and $X_p$ fields both receive the mass  as 
\begin{eqnarray}
		\mathcal{L}_{mass} &=& \frac{1}{2}(\tilde Z^\mu,  \; X_p^\mu) M^2  \left (\begin{array}{c}
			\tilde Z_\mu\\\\
			 X_{p\mu}
		\end{array}
		\right ),
\end{eqnarray}
with the corresponding mass squared matrix element  as
 \begin{eqnarray}\label{mass matrix}
	M^2&=&
	m^2_{\tilde Z}
	\left (\begin{array}{cc}
		1 \;\;\;&\;\;\;s_W \eta\\\\
		s_W \eta\;\;\;&\;\;\; \Delta_Z+ \eta^2  s_W^2
	\end{array}
	\right )\;.
\end{eqnarray}
Here $m_{\tilde Z}^2= g_Z^2 v^2/4$,  $m_{X_p}^2= 4g_D^2v_X^2$ with  $g_Z=\sqrt{g'^2+g^2}$ and $\Delta_Z= m^2_{X_p}/m^2_{\tilde Z}$.

In order to obtain the mass matrix in the physical basis, we need further diagonalize the mass matrix in Eq.~(\ref{mass matrix}) by  introducing the mixing angle as
\begin{eqnarray}\label{mixing}
	&&\left (\begin{array}{c}
		Z\\
		Z'
	\end{array}
	\right )
	=
	\left (\begin{array}{cc}
		c_\theta\;\;\;&\;\;\; s_\theta\\
		-s_\theta\;\;\;&\;\;\;c_\theta
	\end{array}
	\right )
	\left (\begin{array}{c}
		\tilde Z\\
		 X_p
	\end{array}
	\right )\;,\text{with}\nonumber\\
	&& \tan(2\theta) =  \frac{2\eta s_W }{1-\Delta_Z-\eta^2 s_W^2}\approx  \frac{2\epsilon s_W }{1-\Delta_Z}\;.
\end{eqnarray}
The diagonal masses  $ m_Z^2 $ and $ m^2_{Z'}$  are given  by
\begin{eqnarray}
	m^2_Z &=&  m^2_{\tilde Z}  \left[c^2_\theta 
	+ (\Delta_Z+\eta^2 s_W^2)s_\theta^2+  2 s_\theta c_\theta s_W\eta\right]\;,\nonumber\\
 m^2_{Z'} &=&  m^2_{\tilde Z}  \left[s^2_\theta 
	+ (\Delta_Z+\eta^2 s_W^2)c_\theta^2  -2 s_\theta c_\theta s_W\eta\right]\;.
\end{eqnarray}
Due to the assumption $\epsilon<<1$, $m_{Z'}\approx m_{X_p}=2g_D v_X$.

Combining the all transformations in Eqs.~(\ref{caseb}, \ref{SM},\ref{mixing}), we obtain the resulting form as
\begin{eqnarray}
	&&\left (\begin{array}{l}
		 A_p\\\\
		\tilde Z\\\\
	    X
	\end{array}
	\right )
	=\left ( \begin{array}{ccc}
		1\;&\; -s_\theta c_W\eta \;&\;-c_\theta c_W\eta \\\\
		0\;&\; c_\theta+s_\theta s_W\eta  \;&\;\;\; -s_\theta +c_\theta s_W \eta\\\\
		0\;&\;s_\theta\frac{1}{\sqrt{1-\epsilon^2}} \;&\;c_\theta{1\over \sqrt{1-\epsilon^2 }}
	\end{array}
	\right )
	\left (\begin{array}{l}
		A\\\\
		 Z\\\\
		 Z'
	\end{array}
	\right ),
\end{eqnarray}
Here $A_p$ means the original photon component resulting from the mixing of hypercharge field $B$ and  $W^3$ under the condition of $\epsilon=0$.

The general Lagrangian that describes the physical fields ($A$, $Z$, $Z'$)  kinetic energy, and their interactions with the electromagnetic  current $j^\mu_{em}$, neutral $Z$-boson current $j^\mu_Z$ and dark current $j^\mu_X$ is given by
\begin{eqnarray}\label{interaction}
	\mathcal{L} &= &-{1\over 4}  Z'_{\mu\nu}  Z^{\prime \mu\nu} - {1\over 4}  A_{\mu\nu} A^{ \mu\nu} - {1\over 4}  Z_{\mu\nu}  Z^{ \mu\nu} 
	 \nonumber\\
	 && + j^\mu_{em} \left( A_\mu   - s_\theta c_W\eta  Z_\mu -c_\theta c_W\eta Z'_\mu \right)\nonumber\\
	&& + j^\mu_Z\left[ \left( c_\theta+ s_\theta s_W \eta   \right )Z_\mu+ \left(-s_\theta+ c_\theta s_W\eta \right) Z^{\prime}_\mu\right]\nonumber\\
	&&+j^\mu_{Z'}\left(  s_\theta{1\over \sqrt{1-\epsilon^2}}Z_\mu  +c_\theta {1\over \sqrt{1-\epsilon^2  }}  Z'_\mu \right).\label{darkzc}
\end{eqnarray}
 Here the currents for fermions with charge $Q_f$ and weak isospin $I^f_3$ in the SM are given by
\begin{eqnarray}
	&&j^\mu_{em} = - \sum_f  e Q_f \bar f\gamma^\mu f\;,\nonumber\\
	&& j^\mu_Z = -{ e \over 2  s_W  c_W} \bar f\gamma^\mu ( \tilde g^f_V - \tilde g^f_A \gamma_5) f\;,\nonumber\\
    && j^\mu_{Z'}=-i\frac{g_D}{2}(\overline{\chi_2}\gamma^\mu\chi_1-\overline{\chi_1}\gamma^\mu\chi_2)\;.
\end{eqnarray}
here $\tilde g^f_V = I^f_3 - 2 Q_f  s^2_W$ and  $\tilde g^f_A = I^f_3$.

For the linear order approximation in $\epsilon$, $Z'$ interacts with SM fermions as
\begin{eqnarray}\label{Z'_int}
&&\mathcal{L}_{Z^{\prime}\overline{f}f}
= e\epsilon\bar f\gamma^\mu ( c^f_V+ c^f_A \gamma_5) fZ_\mu^{\prime},\\
&& c^f_V=c_W Q_f+ \frac{\tilde g^f_V}{2c_W}
   \frac{\Delta_Z}{1-\Delta_Z}, \;c^f_A=\frac{-\tilde g^f_A}{2c_W}
   \frac{\Delta_Z}{1-\Delta_Z}\;.\nonumber 
\end{eqnarray}
The first term in $c_V$ originates  from the electromagnetic current $j_{em}$, while the remaining terms come from the Z boson current $j_{Z}$. In fact, the terms from $j_{Z}$  play an important role in the production of the new gauge boson $Z'$, especially for large mass $m_{Z'}$ or around $m_Z$, which have usually been neglected in  previous studies~\cite{Filimonova:2022pkj,Lu:2023cet}.

\begin{figure*}[!t]
\centering
 	\subfigure[\label{fig:brx}]
 	{\includegraphics[width=.486\textwidth]{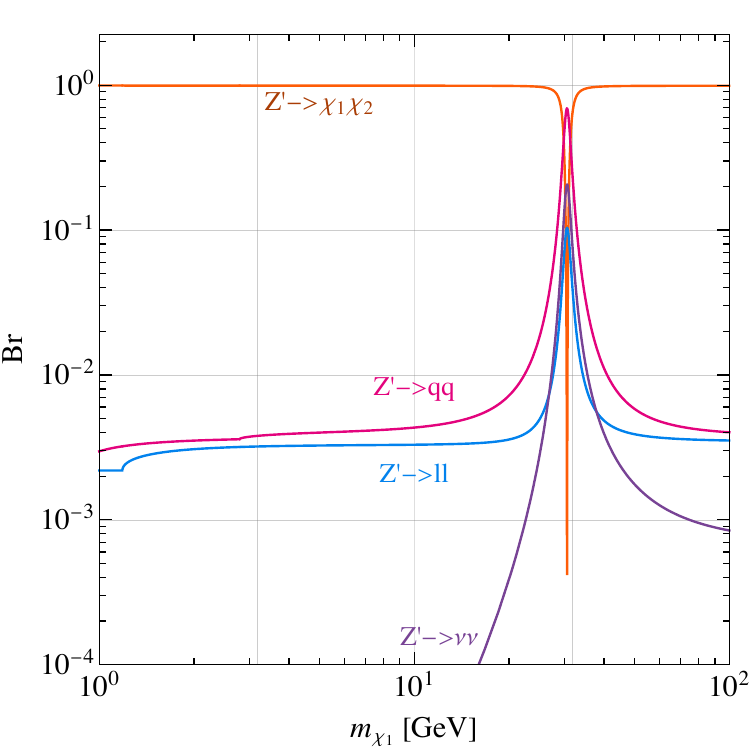}}
 	\subfigure[\label{fig:brchi2}]
 	{\includegraphics[width=.486\textwidth]{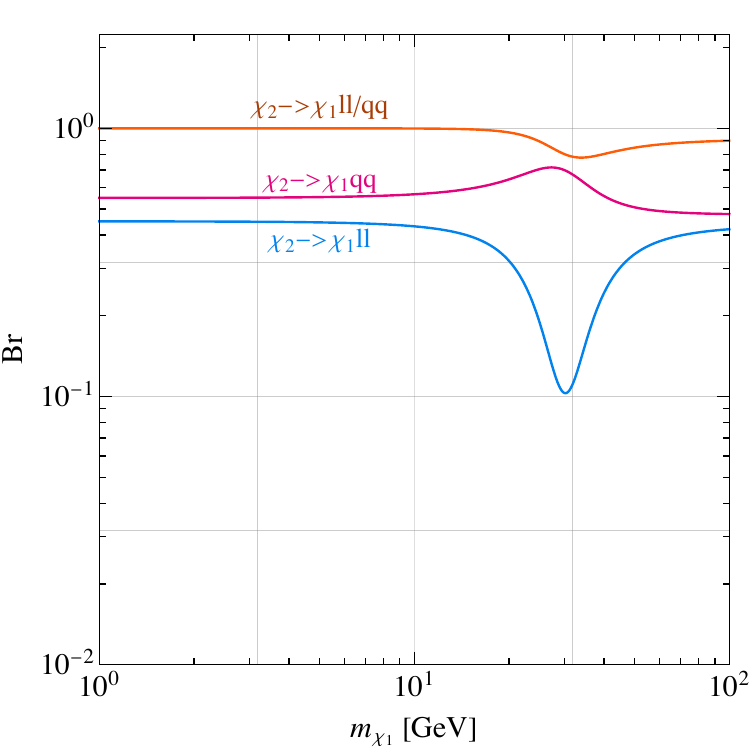}}
	\caption{ The decay branching ratio into different final states for the benchmark case of $m_{\chi_1}=10\Delta_\chi=m_{Z'}/3$,  $\alpha_D=0.1$. 
    The left panel a) shows the  $Z'$ case with  $\epsilon=0.1$.
    The right panel b) means $\chi_2$   decaying into visible SM finals states.
	}
	\label{fig:decay}
\end{figure*}

Before performing the detailed collider analysis, we first analyze the relevant decay processes for the new gauge boson $Z'$ and the excited dark fermion $\chi_2$.
For the gauge boson $Z'$, it can interact with SM fermion in Eq.~(\ref{Z'_int}) and dark fermion in Eq.~(\ref{chi}).
We calculate the corresponding decay width as follows
\begin{eqnarray}\label{Z'decay}
&&\Gamma(Z'\to f\bar f)=\frac{\alpha_{em}\epsilon^2 }{3}m_{Z'}\sqrt{1-\frac{4m_f^2}{m_{Z'}^2}}
    \left[(c_V^f)^2+(c_A^f)^2\right],\nonumber\\
&&\Gamma(Z'\to \chi_1\chi_2)=\frac{m_{Z'}}{12\pi}g_D^2 \lambda\left(1,\; \frac{m_{\chi_1}^2}{m_{Z'}^2},\; \frac{m_{\chi_2}^2}{m_{Z'}^2}\right)\;.
\end{eqnarray}
Here $\lambda(x,y,z)=x^2+y^2+z^2-2xy-2xz-2yz$ and $\alpha_{em}$ is the fine structure constant. $f=q,l,\nu$ represents the SM fermions if the decay is kinematically allowed.
Based on the above formula, we plot the relevant $Z'$ decay branching ratios as shown in Fig.~\ref{fig:brx}.
Here we choose the benchmark points: $m_{\chi_1}=10\Delta_\chi=m_{Z'}/3$,  $\alpha_D=0.1$ and $\epsilon=0.1$. We find that the dominant decay channel is $Br(Z'\to \chi_1\chi_2)\sim 1$. The only exception occurs around $m_{\chi_1}\sim 30$GeV, where it lies at the pole $\Delta_Z\sim 1$ in Eq.~(\ref{Z'_int}), with the enhanced  $Z'$ interactions with SM fermions.  
If choosing  a much smaller kinetic mixing $\epsilon\sim 10^{-3}$, $(Z'\to \chi_2\chi_1)$ becomes the dominant decay chain across all $Z'$ mass regions.

For $\chi_2$, its decay chains are $\chi_2\to \chi_1+\text{SM}$, mediated by $Z'$ via Eqs.~(\ref{chi},\ref{Z'_int}). The SM particles can include a pair of jets, charged leptons, and neutrinos. In the approximation $m_{Z'}>>\Delta_\chi>>m_l$, we can obtain the partial decay width $\chi_2 \to \chi_1 l l$ as
\begin{eqnarray}\label{decay}
    \Gamma(\chi_2\to \chi_1 ll)
    &&=\frac{4\alpha_D \alpha_{em}\Delta_\chi^5\epsilon^2}{9\pi m_{Z'}^4}c_W^2\\
    \times&&\left[
    \left(1-\frac{\tilde g_V^l}{2c^2_W}\frac{\Delta_Z}{1-\Delta_Z}\right)^2+\left(\frac{\tilde g_A^l}{2c_W^2}\frac{\Delta_Z}{1-\Delta_Z}\right)^2
    \right]\nonumber.
\end{eqnarray}
Here $\ell=e,\mu$. We find that  the lifetime of $\chi_2$ depends on $\epsilon$, $\Delta_\chi$ and $m_{Z'}$. Once reducing $\epsilon$ or $\Delta_\chi$, the lifetime will enhance.
The formula can also extended into the final state jet and neutrino scenarios.
Correspondingly, the branching ratios into the visible SM final states are shown  in Fig.~\ref{fig:brchi2}.
Note that the pole around $m_{\chi_1}\sim 30$ GeV is due to the $m_{Z}\sim m_{Z'}$. 
We find that the decay into visible final states (charged leptons and jets) dominates over the neutrino cases.
This provides a viable opportunity to conduct the  collider signal analysis with visible final states.

\section{Lepton Collider Signatures}\label{sec3}

\begin{figure}
    \centering
\includegraphics[width=1.\linewidth]{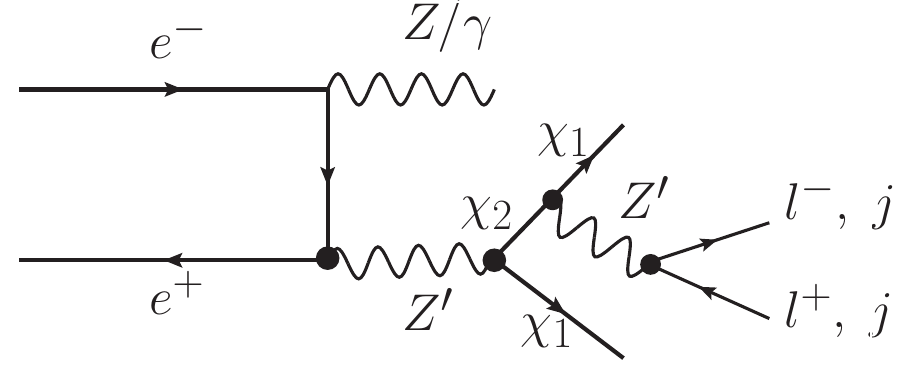}
    \caption{The Feymann diagram for the processes $e^+ e^- \to Z/\gamma + Z', \;Z'\to  \chi_1\chi_2,\; \chi_2\to \chi_1 ll/jj$.}
    \label{fig:feymann}
\end{figure}

In this section, we  discuss the production of inelastic DM at the lepton colliders and classify the signal signatures. 
First of all, the UFO model file of the inelastic
DM model is generated using FeynRules~\cite{Christensen:2008py}, and then we apply MadGraph5 aMC@NLO~\cite{Alwall:2014hca} to
generate Monte Carlo events and calculate cross sections for the following signal process,
\begin{eqnarray}
    e^+e^- \to Z'+ Z/\gamma,\;  Z'\to \chi_2+ \chi_1, \chi_2\to \chi_1 ll/jj\;. 
\end{eqnarray}
The Feymann diagram for relevant process is shown in Fig.~\ref{fig:feymann}. 

In order to conduct the detailed analysis, we adopt the following model parameter benchmarks,
\begin{eqnarray}
    m_{\chi_1}=10\Delta_\chi=m_{Z'}/3, \; \alpha_D=\frac{g_D^2}{4\pi}=0.1\;.
\end{eqnarray}
The values are commonly adopted in many studies,  as cited in Refs.~\cite{Izaguirre:2015zva,Izaguirre:2017bqb,Berlin:2018jbm,Kang:2021oes,Bertuzzo:2022ozu,Lu:2023cet}. Therefore, we use them as benchmark examples to illustrate our analysis.
 Other choices are presented in the Appendix.
The $\chi_1$ mass can vary in the range below,
\begin{eqnarray}
    m_{\chi_1}/\text{GeV}=(1,\;100)\;.
\end{eqnarray}
This is consistent with the observed DM relic density~\cite{Izaguirre:2015zva,Izaguirre:2017bqb,Duerr:2019dmv} for $\chi_1$ as the DM candidate.

Before performing a detailed collider analysis, we first examine the corresponding constraints on the model parameters.
\begin{itemize}
\item $(g-2)_\mu$: exchanging $Z'$ at one loop level gives the contribution as 
\begin{eqnarray}
    \Delta a_\mu&=&\frac{\alpha_{em}\epsilon^2m_\mu^2}{\pi m^2_{Z'}}\int_0^1dx \frac{1}{1-x+\left(m_{\mu}/m_{Z'}\right)^2x^2}
    \\
    &\times &\left[(c_V^\mu)^2x^2(1-x)+(c_A^\mu)^2(x-x^2)(x-4)\right]\;,\nonumber
\end{eqnarray}
The most recent measurement of $(g-2)_\mu$ was  performed by the Fermilab experiment~\cite{Muong-2:2023cdq}, showing a deviation from the recent SM prediction~\cite{Aliberti:2025beg}, $a_\mu=(26\pm66)\times 10^{-11}$.
We choose 2$\sigma$ errors to place an exclusive limit  on the model parameters $\epsilon-m_{Z'}$ in the   upper left corner of Fig.~\ref{fig:bound}. 
\item \textbf{electroweak fit}: the effects are parameterized by the oblique parameters (STU)~\cite{Curtin:2014cca,Harigaya:2023uhg}, given by
\begin{eqnarray}
&&\alpha S=4\epsilon^2 s_W^2 c_W^2 \frac{1}{1-\Delta_Z}\left(1-\frac{s_W^2}{1-\Delta_Z}\right)\;,\nonumber\\
&& \alpha T=-\frac{\epsilon^2 s_W^2}{(1-\Delta_Z)^2}\Delta_Z \;,\quad\alpha U=\frac{4\epsilon^2 s_W^4 c_W^2}{(1-\Delta_Z)^2}\;.
\end{eqnarray}
The current electroweak fit requires $S=-0.04\pm0.10$, $T=0.01\pm 0.12$ and $U=-0.01\pm 0.09$~\cite{ParticleDataGroup:2024cfk}, which can give the upper bound on $\epsilon<6\times 10^{-3}$ around Z boson mass in Fig.~\ref{fig:bound}.
\item \textbf{collider bounds}: other experimental constraints derive from LEP~\cite{Hook:2010tw,Curtin:2014cca}, BaBar~\cite{BaBar:2017tiz}, and CMS~\cite{CMS:2023bay}, respectively, as shown in the gray regions of Fig.~\ref{fig:bound}.
\item \textbf{DM abundance}: the $\chi_1$ abundance matches the observed DM relic density~\cite{Izaguirre:2015zva,Izaguirre:2017bqb,Duerr:2019dmv}, which is indicated in the black line. 
 A dimensionless parameter $y$ is defined as
\begin{eqnarray}\label{yy}
   y= \epsilon^2 \alpha_D\frac{m^4_{\chi_1}}{m_{Z'}^4}\;.
\end{eqnarray}
It uniquely determining the freeze-out annihilation rate by a given parameter combinations, frequently adopted in DM abundance analysis~\cite{Izaguirre:2015zva,Izaguirre:2017bqb,Berlin:2018jbm}. 
The details can refer to  Appendix.
\end{itemize}

Now we conduct the detailed collider analysis.
The $Z'$ production occurs via $e^+e^-\to Z/\gamma+Z'$, characterized by kinetic mixing parameter $\epsilon$ in Eq.~(\ref{Z'_int}). Ignoring the masses of initial electron and positron, we approximately calculate the corresponding cross sections as
\begin{eqnarray}\label{production}
&& \frac{\sigma(e^+e^-\to Z'\gamma)}{\sigma(e^+e^-\to Z\gamma)}\approx 4\epsilon^2s_W^2 c_W^2
    \frac{(c_V^e)^2+(c_A^e)^2}{(\tilde g_V^e)^2+\tilde g_A^e)^2}\;,\\
&&\frac{\sigma(e^+e^-\to Z'Z)}{\sigma(e^+e^-\to ZZ)}\approx 8\epsilon^2s_W^2 c_W^2
    \frac{(c_V^e\tilde g_V^e)^2+(c_A^e\tilde g_A^e)^2}{(\tilde g_V^e)^4+\tilde g_A^e)^4}\;.\nonumber
\end{eqnarray}
Note that the additional factor of 2 in second line comes from the final states of identical $Z$ bosons in $e^+e^-\to ZZ$.
The subsequent decays of $Z'$ and $\chi_2$ are shown in
Fig.~\ref{fig:decay}.

\begin{table*}[htbp!]
\centering
\caption{ The relevant information for different lepton ($e^+e^-$) colliders, including CLIC~\cite{Aicheler:2018arh}, ILC~\cite{ILD:2019kmq},
FCC-ee~\cite{FCC:2018evy} and CEPC~\cite{CEPCStudyGroup:2023quu},   
includes
expected center-of-mass energy,   integrated luminosity  and detector paramaters. 
The corresponding  profile sketch of the detector components  includes  the inner tracker,  HCAL and muon chamber (with barrel and endcaps) as shown in Ref.~\cite{Cheung:2019qdr}.
Here $R_e$ means incoming radius in endcaps.
}
\begin{tabular}{|c|c|c|c|c|c|c|c|c|c|c|c|c|c|c|c}
\hline \multirow{2}{*}{Facility} & \multirow{2}{*}{$\sqrt{s}(\mathrm{GeV})$} & \multirow{2}{*}{$\mathcal{L}_{lum}(\mathrm{ab}^{-1})$} 
& \multicolumn{3}{|c|}{inner tracker} & \multicolumn{5}{|c|}{muon system} & 
\multicolumn{4}{|c|}{HCAL} \tabularnewline\cline{4-15}  
& & & $R_I$(mm) & $R_{o}$(m) & $\pm z$(m) & $R_e(m)$&  $R_I$(mm) & $R_{o}$(m) & $z_I$(m) & $z_o$(m)& $R_I$(mm) & $R_{o}$(m) & $z_I$(m)
& $z_o$(m) \\\hline
ILC & $250 $ & $2 $ 
 &
16 & 1.776 & 2.212 & 0.58&
4.45 & 7.755 & 4.072 & 6.712 & 2.058 & 3.345 & 2.65 & 3.937\\
\hline
FCC-ee & \begin{tabular}{c}
$91 $ \\
$160 $ \\
$240 $ 
\end{tabular} & \begin{tabular}{c}
$150 $ \\
$10 $ \\
$5 $ 
\end{tabular}  &
17 & 2 & 2 & 0.35
&  4.5 & 5.5& 5.5 & 6.5 & 2.5  & 4.5 &  3 & 5.5  \\
\hline  CEPC & \begin{tabular}{c}
$91 $ \\
$160$ \\
$240 $ \\
\end{tabular} & \begin{tabular}{c}
$100 $ \\
$6.9 $ \\
$21.6 $ \\
\end{tabular}&
16 & 1.81 & 2.35 & 0.5
& 4.4 & 6.08 &4.14 & 5.86& 2.058 & 3.144 & 2.65 & 3.736 \\
\hline
\end{tabular}
\label{Collider}
\end{table*}

In order to facilitate a more detailed analysis of collider signals, we firstly elaborate on the different types of lepton colliders, particularly the electron-positron scenario.
The particle physics community has actively discussed the possibility of building a new lepton collider to reduce uncertainties and improve precision measurements.
Currently, there are two types of proposals for the $e^+e^-$ case: circular and linear colliders as follows.
\begin{itemize}
\item \textbf{CLIC}: the Compact Linear Collider~\cite{Aicheler:2018arh}, with a length ranging from 11 to 50 km and a proposed energy range of (380, 1500, 3000) GeV.
\item \textbf{ILC}: the International Linear Collider~\cite{Barklow:2015tja,ILD:2019kmq}, 
with a length of approximately 20 km and proposed energies of (250, 365, 500) GeV.
\item \textbf{FCC-ee}: the Future Circular Collider~\cite{FCC:2018evy},  
with a 90 km circular ring and collider energies of (91, 160, 240, 365) GeV.
\item \textbf{CEPC}: the Circular Electron-Positron Collider~\cite{CEPCStudyGroup:2023quu,Ai:2024nmn}, with a 100 km circumference and the same energies as the FCC-ee, albeit with smaller luminosities.
\end{itemize}

Based on the above description and the range of $m_{\chi_1}$, we primarily study the collider signatures for energies below approximately 300 GeV. Furthermore, we focus on investigating the collider signatures for ILC (250 GeV), FCC-ee (91 GeV, 160 GeV, 240 GeV), and CEPC (91 GeV, 160 GeV, 240 GeV), respectively.
Here, the numbers in parentheses represent the collider center-of-mass energies.
Correspondingly, the proposed center-of-mass energies, luminosities, and geometrical structures are summarized in Table.~\ref{Collider}.
The profile sketch of the detector components of lepton colliders can refer to Fig. 3 in Ref.~\cite{Cheung:2019qdr}.

Since the final states from $\chi_2$ decays involve charged leptons or jets, we will analyze the corresponding collider signatures both individually and in combination.
The total signal events $N_{s.e.}$ can be obtained by reconstructing displaced vertices in the above regions using Monte Carlo simulations, as given by
\begin{eqnarray}
N_{\mathrm{s.e.}}
&=&\mathcal{L}\cdot\sigma\cdot\frac{1}
{N^{MC}}\sum_{i=1}^{N^{MC}}P_i^{DV}\epsilon^i_{DV}\;. 
\end{eqnarray}
Here, $\mathcal{L}$ represents the integrated luminosity of the corresponding lepton collider, and $\sigma$ denotes the cross section for the process $e^+ e^-\to \gamma/Z +Z',\;Z'\to  \chi_1\chi_2,\; \chi_2\to \chi_1+ \text{vis}$. 
$N^{MC}$ is the total number of MC-simulated events.
$P_i^{DS}$ means the individual decay probability of the long-lived particle $\chi_2$ in the $i$-th simulated signal event within the respective fiducial component. 
$\epsilon_{DV}$ represents the efficiency of our displaced vertex analysis, which depicts the reconstruction effects, with $\epsilon_{DV} = 100\%$ for simplicity unless otherwise stated~\cite{Liu:2022kid}.
For the details of the Monte Carlo simulations, we use inverse sampling of the decay distribution to simulate the displacement.

As shown in Table.~\ref{Collider}, we consider three distinct detector regions: the interaction point (IP), the muon system (MS), and the hadronic calorimeter (HCAL).
Note that the final visible particles from $\chi_2$ decays could either be a pair of charged leptons or jets.
In the former case (charged leptons), we consider both the IP and the MS for reconstructing the displaced vertices. In contrast, for jets, we additionally include the  HCAL as part of the fiducial volume.
Furthermore, the vertex detector and inner tracker for the IP are considered together.
 The HCAL and MS consist of a barrel and two endcaps.

Regardless of which of the aforementioned three regions the particle decay occurs in, the visible final states can be detected.
Taking into account the Lorentz boost effects, the decay width of $\chi_2$ is
\begin{eqnarray}\label{LLP}
    \lambda=\beta_i \gamma_i \tau_{\chi_2}=\frac{p_i}{E_i}\frac{E_i}{m_{\chi_2}}\tau_{\chi_2}
    =\frac{p_i}{m_{\chi_2}}\tau_{\chi_2}\;,
\end{eqnarray}
here $\beta_i(\gamma_i)$ means the speed (boost) factor of $\chi_2$, and $\tau_{\chi_2}$ denotes its lifetime.
By defining the polar angle between the traveling direction of $\chi_2$ and the z-axis as $\theta_i$, we can derive the corresponding selection criteria for observing $\chi_2$ decays within the aforementioned three regions as follows:
\begin{itemize}
    \item \textbf{IT}: $R_I<|\lambda_i\sin\theta_i|<R_o$ and $|\lambda_i\cos\theta_i|<z$.
     \item \textbf{Barrel}: $R_I<|\lambda_i\sin\theta_i|<R_o$ and $|\lambda_i\cos\theta_i|<z_I$.
    \item  \textbf{Endcaps}: $R_e<|\lambda_i\sin\theta_i|<R_o$ and $z_I<|\lambda_i\cos\theta_i|<z_o$.
\end{itemize}
Regardless of the regions in which the decays occur, the visible final state signals can be detected.
Furthermore, to obtain reliable results, we impose the following cuts:
\begin{eqnarray}\label{cut}
    p_{tl}>2.5\text{GeV},\; \quad p_{tj}>5\text{GeV}\;.
\end{eqnarray}
This corresponds to the requirement that $\Delta_\chi>5 (10)$ GeV for $ll(jj)$ final states.

\begin{figure*}[!t]
\centering
 	\subfigure[\label{cross}]
 	{\includegraphics[width=.486\textwidth]{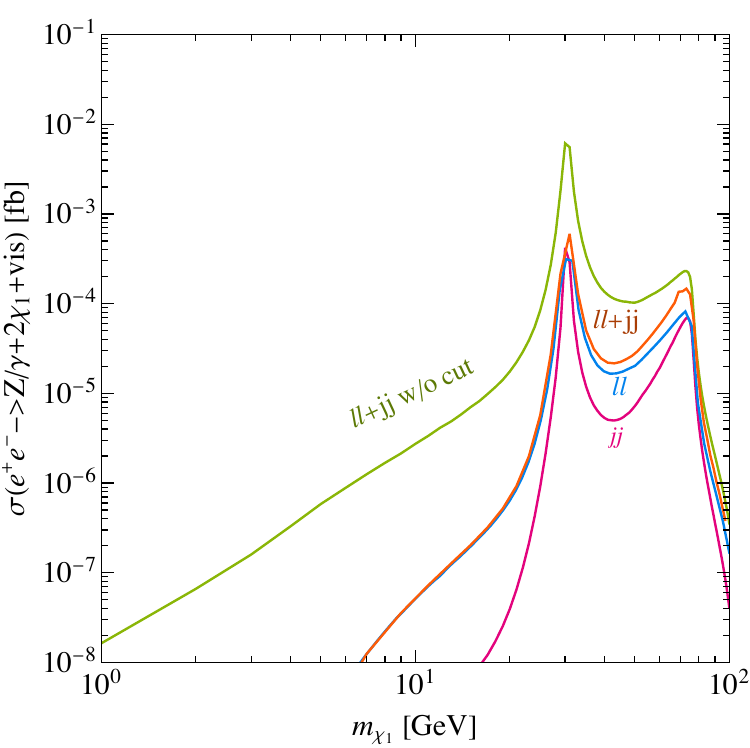}}
 	\subfigure[\label{cepc}]
 	{\includegraphics[width=.486\textwidth]{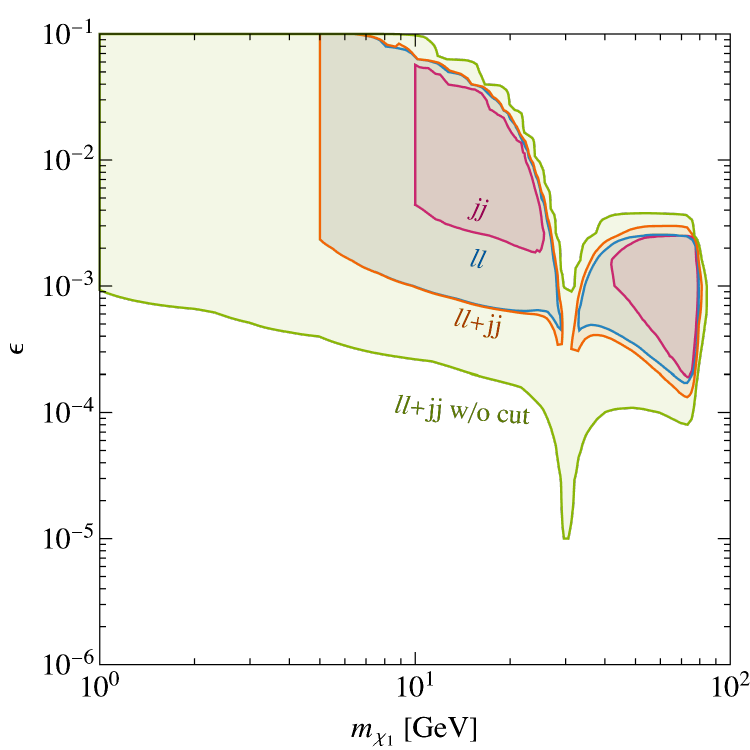}}
	\caption{ 
   The cross section  $\sigma(e^+e^-\to Z/\gamma+2\chi_1+ll/qq)$ and sensitivity on the model parameters for the 240 GeV CEPC.   Here we choose the benchmark $m_{\chi_1}=10\Delta_\chi=m_{Z'}/3$,  $\alpha_D=0.1$. 
  The green represents the visible final states ($ll+jj$) without the cut. The remaining ones correspond to the cases by imposing cut in Eq.~(\ref{cut}), with magenta for $jj$, blue for $ll$ and red for $ll+jj$, respectively.
    The left panel a) shows the  $m_{\chi_1}-\sigma$ plane for fixing  $\epsilon=10^{-4}$.
    The right panel b) means the  $m_{\chi_1}-\epsilon$ plane for different scenarios.
	}
	\label{fig:cepc240}
\end{figure*}

For convenience, we use the CEPC (240 GeV) as an example to illustrate the analysis details.
The relevant results are shown in Fig.~\ref{fig:cepc240}. Here we choose the benchmark $m_{\chi_1}=10\Delta_\chi=m_{Z'}/3$,  $\alpha_D=0.1$. 
Four different scenarios are analyzed, marked with different colors, depending on whether the cuts in Eq.~\ref{cut} are imposed and the final states (charged leptons or jets).
In Fig.~\ref{cross}, the cross section $\sigma(e^+e^-\to Z/\gamma+2\chi_1+ll/qq)$  is obtained for $\epsilon=10^{-4}$ in terms of the DM $\chi_1$ mass $m_{\chi_1}$.
We find two distinct peaks around $m_{\chi_1}\sim 30$ GeV and $m_{\chi_1}\sim 80$ GeV. 
The former case corresponds to $m_{Z'}\sim m_{Z}\sim 90$ GeV, arising from the interaction coupling pole in Eq.~(\ref{Z'_int}).
The latter is due to the contribution from $e^+e^- \to \gamma Z'$ induced by $m_{Z'} \sim \sqrt{s}$.
Moreover, the cross section for visible final states ($ll + jj$) without the cuts can reach a maximum of 0.01 fb around $m_{\chi_1} \sim 30$ GeV, as indicated in green.
After imposing the cuts in Eq.~(\ref{cut}), the corresponding maximum cross section decreases by one order of magnitude to 0.001 fb in red, followed by $ll$ in blue and $jj$ in magenta.
 And the dominant contribution comes from the charged lepton final states.
 The corresponding event numbers after imposing the cuts are shown in Tab.~\ref{number}.

 \begin{table*}[htbp!]
\centering
\caption{ The signal event numbers for different lepton colliders after imposing the cuts. Here we choose the benchmark $m_{\chi_1}=10\Delta_\chi=m_{Z'}/3$,  $\alpha_D=0.1$. 
}
\begin{tabular}{|c|c|c|c|c|c|c|c|c|c|c|c|c|}
\hline \multirow{3}{*}{Facility} & \multirow{3}{*}{$\sqrt{s}(\mathrm{GeV})$} & \multirow{3}{*}{$\mathcal{L}_{lum}(\mathrm{ab}^{-1})$} 
& \multicolumn{6}{|c|}{event numbers}\tabularnewline\cline{4-9} 
&&& \multicolumn{3}{|c|}{$\epsilon=10^{-3}$} & \multicolumn{3}{|c|}{$\epsilon=10^{-2}$} \tabularnewline\cline{4-9}  
& & & $m_{\chi_1}=10$GeV & $m_{\chi_1}=40$GeV & $m_{\chi_1}=70$GeV&  $m_{\chi_1}=6$GeV & $m_{\chi_1}=10$GeV & $m_{\chi_1}=20$GeV \\\hline
ILC & $250 $ & $2 $ 
 &0.34  & 0.61 & 3.12 &21.40 & 56.81 & 13.89 \\
\hline
FCC-ee & \begin{tabular}{c}
$91 $ \\
$160 $ \\
$240 $ 
\end{tabular} & \begin{tabular}{c}
$150 $ \\
$10 $ \\
$5 $ 
\end{tabular}  &
\begin{tabular}{c}
1.77 \\
1.07 \\
0.65 \\
\end{tabular}&
\begin{tabular}{c}
0 \\
5.65\\
1.49 \\
\end{tabular}&
\begin{tabular}{c}
0 \\
 0\\
8.44 \\
\end{tabular}&
\begin{tabular}{c}
79.70 \\
 81.19\\
53.83\\
\end{tabular}&
\begin{tabular}{c}
202.22 \\
176.29 \\
139.39 \\
\end{tabular}&
\begin{tabular}{c}
23.80 \\
24.43 \\
29.41 \\
\end{tabular}
        \\
\hline  CEPC & \begin{tabular}{c}
$91 $ \\
$160$ \\
$240 $ \\
\end{tabular} & \begin{tabular}{c}
$100 $ \\
$6.9 $ \\
$21.6 $ \\
\end{tabular}&
\begin{tabular}{c}
1.19 \\
0.78 \\
 3.10\\
\end{tabular}&
\begin{tabular}{c}
0 \\
4.18 \\
6.79\\
\end{tabular}&
\begin{tabular}{c}
0 \\
0 \\
38.66 \\
\end{tabular}&
\begin{tabular}{c}
53.66 \\
56.45 \\
234.02 \\
\end{tabular}&
\begin{tabular}{c}
138.99 \\
 124.27\\
612.88 \\
\end{tabular}&
\begin{tabular}{c}
20.25 \\
19.64\\
142.90 \\
\end{tabular}   \\
\hline
\end{tabular}
\label{number}
\end{table*}

\section{Sensitivity}\label{sec4}

In this section, we discuss the sensitivity of the lepton collider to the model parameters $m_{\chi_1}$ and $\epsilon$.

As shown in Fig.~\ref{cepc}, the sensitivity to the model parameters is analyzed for four distinct scenarios, depicted by different colors.
We find that the scenario of $ll + jj$ without cuts, marked by the green region, exhibits excellent sensitivity with $O(10^{-3})$, even approaching $\epsilon \sim 10^{-5}$ around $m_{\chi_1}\sim 30$ GeV.
Additionally, this scenario ($ll + jj$ without cuts) covers the entire $\chi_1$ mass range.
After imposing the cut, the sensitivity region induced by the same final states narrows down and reduces by one order of magnitude for $\epsilon$.
Furthermore, the individual scenarios for charged leptons or jets are shown separately.
The sensitivity from the charged lepton scenario is significantly superior to that from the jet scenarios, indicating that the dominant contribution to the $ll + jj$ final states comes from the charged lepton case.
Therefore, we mainly focus on the case of charged lepton final states for different types of lepton colliders in the following.

\begin{figure*}
    \centering
    \includegraphics[width=1.\linewidth]{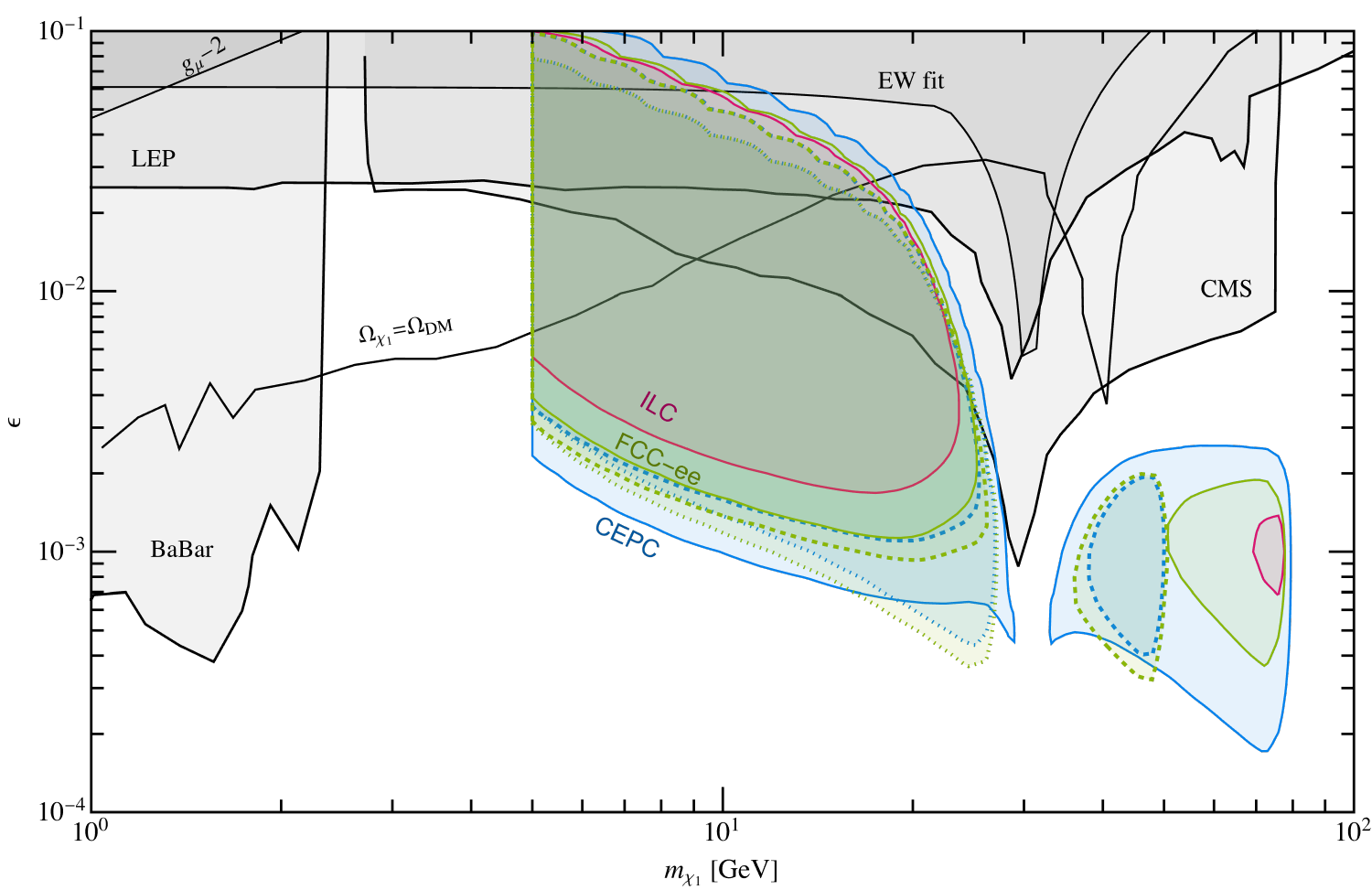}
    \caption{The current experimental bounds and future lepton collider sensitivities in  the  $m_{\chi_1}-\epsilon$ plane.
    Here we choose the benchmark $m_{\chi_1}=10\Delta_\chi=m_{Z'}/3$ and  $\alpha_D=0.1$.
    The light gray regions represent the excluded regions from $(g-2)_\mu$~\cite{Muong-2:2023cdq,Borsanyi:2020mff}, electroweak fit~\cite{ParticleDataGroup:2024cfk,Harigaya:2023uhg}, LEP~\cite{Hook:2010tw,Curtin:2014cca}, BaBar~\cite{BaBar:2017tiz}, and CMS~\cite{CMS:2023bay}, respectively.
    The black line represents the region where the abundance of $\chi_1$  matches the observed DM relic density~\cite{Izaguirre:2015zva,Izaguirre:2017bqb,Duerr:2019dmv}.
    Different types of lepton colliders are shown in distinct colors, with the
    corresponding center-of-mass energy 240/250 GeV~(solid), 160 GeV~(dashed), 91 GeV~(dotted). }
    \label{fig:bound}
\end{figure*}

The relevant findings are illustrated in Fig.~\ref{fig:bound}, which include both current bounds in light gray regions and the projected lepton collider sensitivity in colored regions.
The gray regions  summarize the current experimental constraints as mentioned before.

The colored regions represent the projected sensitivity from different lepton colliders.
Our analysis demonstrates that all lepton colliders are capable of probing new parameter regions that are not excluded by current bounds, especially the $\chi_1$ mass regions in $(5 - 28)$ GeV and $(32 - 80)$ GeV.
Two distinct features are observed, as already mentioned in the previous   cross section in Fig.~\ref{cross}.
First, a kinematic separation emerges near $m_{\chi_1}\sim 30$ GeV, where $\chi_2$'s short decay lifetime ($\tau_{\chi_2}$) precludes the possibility of decay within the IT, MS, or HCAL subsystems.
When increasing $\tau_{\chi_2}$ by reducing $\epsilon$, the $Z'$ production rate scales as $\epsilon^2$, leading to significant suppression.
Second, an upper mass limit $m_{\chi_1}<80$ GeV appears when
$\sqrt{s}\sim m_{Z'}$, reflecting the kinematic threshold.
Besides, distinct performance characteristics emerge across lepton colliders.
For CEPC, 240 GeV achieves optimal sensitivity, exceeding 90 GeV and 160 GeV performance,  owing to maximal beam energy and enhanced luminosity.
And CEPC 91 GeV  demonstrates unique sensitivity in the (18 - 26) GeV mass range, whereas the 160 GeV operation exhibits reduced sensitivity primarily due to luminosity constraints.
On the other hand,  while operating at same energies, FCC-ee displays marginally different sensitivity profiles due to variations in luminosity and detector acceptance.
Each collider energy  probes distinct parameter space regions with characteristic sensitivity patterns.
Notably, FCC-ee 91 GeV outperforms its 160 GeV  and 240 GeV, attributable to superior luminosity conditions. 
We also find that
 ILC (250 GeV) has comparatively weaker sensitivity in the majority of intervals, with its coverage fully encompassed by CEPC and FCC-ee at 240 GeV.
Combining these projected sensitivities from different lepton colliders, we find that it can approach $\epsilon\sim 2\times 10^{-4}$ at best, which is clearly advantageous compared to the results from LHC~\cite{Berlin:2018jbm,Bertuzzo:2022ozu,Lu:2023cet} and Belle~\cite{Duerr:2019dmv,Duerr:2020muu,Kang:2021oes}.

\section{conclusion}\label{sec5}

In this work, we investigate the detection prospects for the inelastic dark model with an additional  $U(1)_D$ gauge symmetry at various lepton colliders.
The new gauge boson $Z'$ from  $U(1)_D$ serves as a portal to connect the SM fermions and DM fermions $\chi$.
The kinetic mixing $\epsilon$ between $Z'$  and the hypercharge field $B$ leads to  the $Z'$  production.
And the $Z'$  interaction with dark fermions $\chi_{1,2}$ further induces the decay chain $Z'\to \chi_1\chi_2$. Here $\chi_{1}$ ($\chi_{2}$) represents the ground (excited) DM fermion, 
with $\chi_1$ acting as the DM candidate and $\chi_2$ as a long-lived particle respectively.
In order to saturate the observed DM relic density,
the mass difference $\Delta_\chi=\chi_2-\chi_1$, induced by the interaction between the dark Higgs field and the DM sector, needs to be small enough, leading to the subsequent decays $\chi_2\to \chi_1 +\text{vis}$.
Based on the sensitivity of Belle and the LHC from previous studies, we explore the projected sensitivity from future lepton colliders, which benefit from cleaner environments and higher luminosities.
For the inelastic DM mass $m_{\chi_1}$ in the range of (1-100) GeV, we find that the future lepton colliders can  probe many unexplored regions beyond current experimental bounds and provide better projected sensitivity than the LHC and Belle, notably for CEPC 240 GeV approaching $\epsilon\sim 10^{-5}$ without cuts.
 This indicates that the future collider can serve as a complementary tool for exploring the inelastic dark matter model.
\\

\section*{Acknowledgments}

W.L. is supported by National Natural Science foundation of China (Grant
No. 12205153).
J. S. was supported by IBS under the project code, IBS-R018-D1.

\section*{Appendix: DM abundance}

In the standard thermal freeze-out scenario, the DM abundance today is fixed by the DM number density at the time when its annihilation rate into Standard Model (SM) states drops below the Hubble expansion rate.
The DM $\chi_1$ relic abundance is governed by  Boltzmann equations as
\begin{eqnarray}
  \frac{dY_{1,2}}{dx}&=&-\frac{\lambda_A}{x^2}\left(Y_1Y_2-Y_1^{(0)}Y_2^{(0)}\right)\pm x\lambda_D\left(Y_2-\frac{Y_2^{(0)}}{Y_1^{(0)}}Y_1\right)\nonumber\\
  &\pm&\frac{\lambda_S}{x^2}Y_f^{(0)}\left(Y_2-\frac{Y_2^{(0)}}{Y_1^{(0)}}Y_1\right),  
\end{eqnarray}
here $x=m_{\chi_2}/T$ means a dimensionless time variable, and $Y_i=n_{\chi_i}/s(T)$ is  comoving number density normalized to the entropy density $s(T)=2\pi^2g_{s,*}T^3/45$ with 
the number of entropic degrees of freedom $g_{s,*}$.
The superscript (0)  denotes an equilibrium quantity. $\lambda_{A,S,D}$ are the collision terms from thermally averaged
coannihilation, inelastic scattering and  decay processes, respectively, defined by
\begin{eqnarray}
   && \lambda_{A}=\frac{s(m_{\chi_2})}{H(m_{\chi_2})}\langle\sigma v(\chi_1\chi_2\to f\bar{f})\rangle\;,\nonumber\\
    &&\lambda_{S}=\frac{s(m_{\chi_2})}{H(m_{\chi_2})}\langle\sigma v(\chi_2f\to\chi_1f)\rangle\;,\nonumber\\
&&\lambda_{D}=\frac{\langle\Gamma(\chi_2\to\chi_1f\bar{f})\rangle}{H(m_{\chi_2})}\;.
\end{eqnarray}
Here the Hubble rate $H(T)\approx 1.66\sqrt{g_*}T^2/M_{pl}$, where $g_*$ is the number of relativistic
degrees of freedom and the Planck
mass $M_{pl}=1.2\times 10^{19}$ GeV.
The decay rates $\Gamma(\chi_2\to \chi_1 f\bar f)$ are indicated in Eq.~(\ref{decay}).
And the relevant cross sections  for the thermal relic abundances are 
\begin{eqnarray}
&& \langle\sigma v(\chi_1\chi_2\to\bar{f}f)\rangle=\frac{\int_{s_0}^\infty ds\sigma(s)(s-s_0)\sqrt{s}K_1\left(\frac{\sqrt{s}}{T}\right)}{8m_{\chi_1}^2m_{\chi_2}^2TK_2\left(\frac{m_{\chi_1}}{T}\right)K_2\left(\frac{m_{\chi_2}}{T}\right)}\;,\nonumber\\
&&\langle\sigma v(\chi_2e\to\chi_1e)\rangle\sim 16\pi\epsilon^2c_W^2 \alpha_{em}\alpha_D\frac{T^2}{m_{Z'}^4}\;,
\end{eqnarray}
here $s_0=(m_{\chi_1}+m_{\chi_2})^2$, $K_1$ and $K_2$ are the modified Bessel functions of the first and second kind. Additionally, in the leading order,   the corresponding
s-wave co-annihilation cross section can be written as
\begin{eqnarray}
\sigma(\chi_1\chi_2\to\bar{f}f)&\approx&\frac{4\pi\epsilon^2c_W^2\alpha_{em}\alpha_D(m_{\chi_1}+m_{\chi_2})^2}{[(m_{\chi_1}+m_{\chi_2})^2-m_{A^{\prime}}^2]^2+m_{Z'}^2\Gamma_{Z'}^2}\nonumber\\
&\propto & \epsilon^2 \alpha_D \frac{m^4_{\chi_1}}{m_{Z'}^4}\frac{1}{m_{\chi_1}^2}=\frac{y}{m_{\chi_1}^2}
\end{eqnarray}
In the second line we use the limit of $m_{Z'}>>m_{\chi_{1,2}}$.
For our case of $m_{Z'}>m_{\chi_{1,2}}$, then freeze-out
predominantly proceeds through coannihilations
$\chi_1\chi_2\to Z^{\prime *}\to f\bar f$.

Solving this Boltzmann equations, we obtain  $Y_1$ at freeze-out near $x\sim 20$, which determines the relic abundance as
\begin{eqnarray}
  \Omega_{\mathrm{DM}}=\frac{\rho_{\chi_1}}{\rho_{\mathrm{cr}}}=\frac{m_{\chi_1}s_0Y_1}{\rho_{\mathrm{cr}}}=0.12 h^2;,  
\end{eqnarray}
where the critical density is $\rho_{\mathrm{cr}}=8.1h^2\times 10^{-47}$GeV$^4$ and the present day CMB entropy $s_0\approx 2969$cm$^{-3}$. Therefore, the relevant DM relic density is shown by the black line in Fig.~\ref{fig:bound}.

\begin{figure*}[!t]
\centering
\subfigure[\label{bound25}]
 {\includegraphics[width=.486\textwidth]{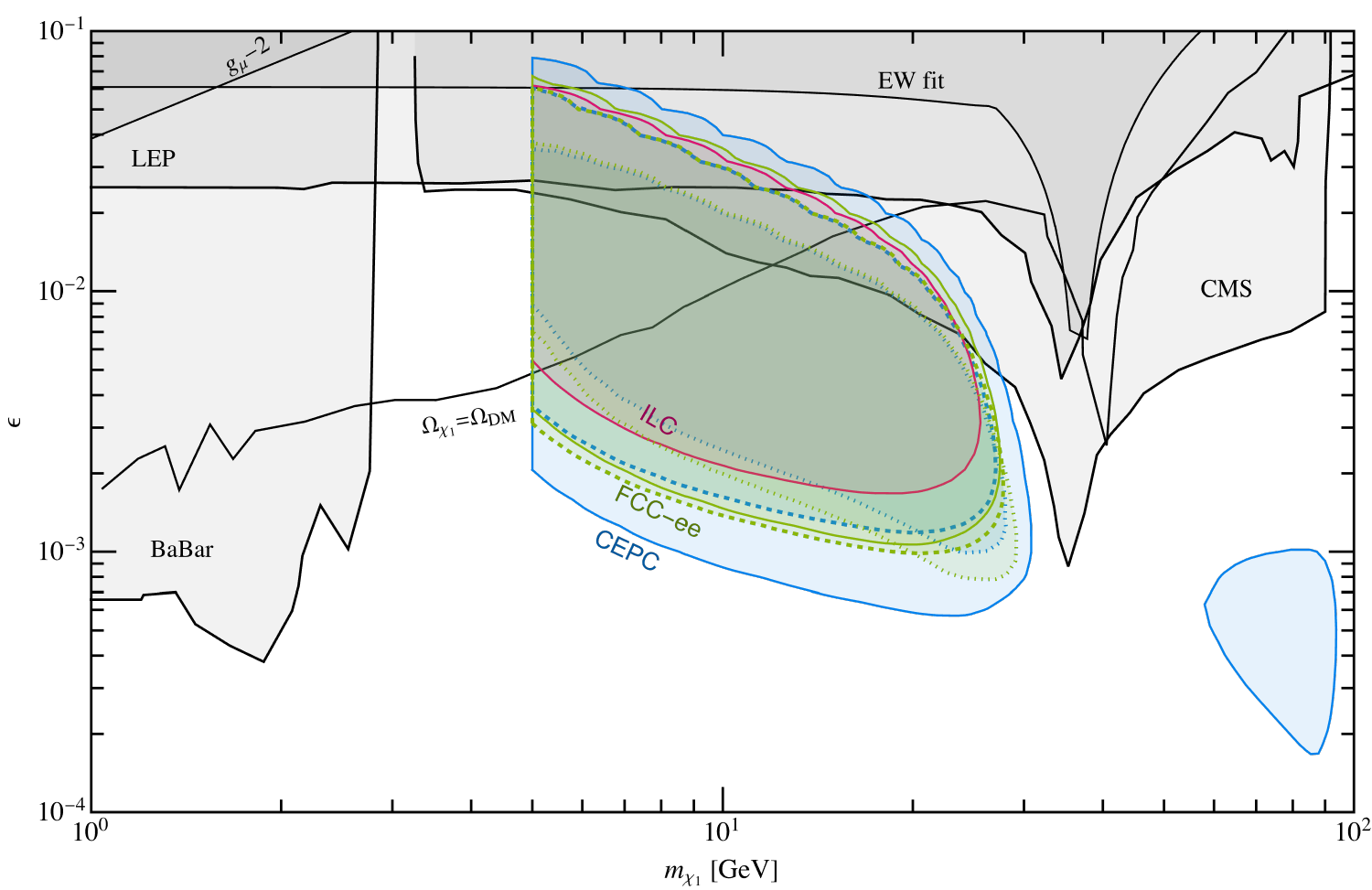}}
 	\subfigure[\label{bound6}]
 {\includegraphics[width=.486\textwidth]{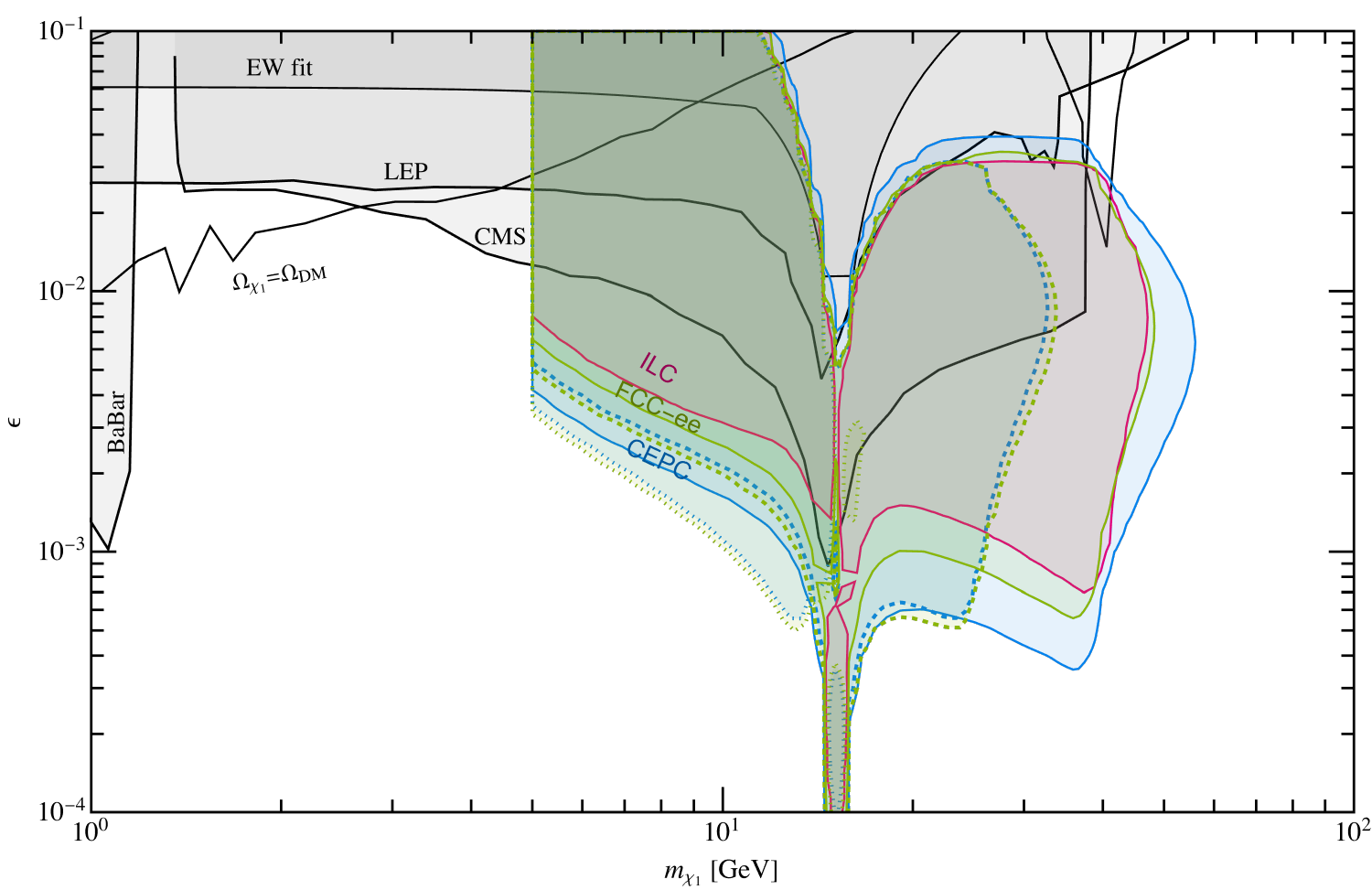}}
	\caption{ 
   The same as Fig.~\ref{fig:bound} for the left panel $m_{Z'}/m_{\chi_1}=2.5$ and the right panel  $m_{Z'}/m_{\chi_1}=6$.
	}
	\label{fig:ratio}
\end{figure*}

The corresponding constraints for alternative values of the mass ratio $m_{Z'}/m_{\chi_1}$  are shown in Fig.~\ref{fig:ratio}, with $m_{Z'}/m_{\chi_1}=2.5$(10) in the left (right) panel.
As the ratio $m_{Z'}/m_{\chi_1}$  increases, the  current bounds shift leftward, including  LEP, BaBar, CMS, $(g-2)_\mu$ and  EW fit, as their bounds are all  dependent on the $Z'$ mass.
We find that  as the ratio $m_{Z'}/m_{\chi_1}$ increases, both the relic density curve
and projected sensitivities line   shift upward. 
This shift arises from the correlation between the dark matter relic density and collider signatures.
Considering the production and decay processes described in Eqs.(\ref{production},\ref{Z'decay},\ref{decay}),  and the long-lived particle $\chi_2$ in Eq.~(\ref{LLP}), the total event rate scales approximately as $\epsilon^4 \alpha_D m^4_{\chi_1}/m_{Z'}^4=\epsilon^2 y$, where $\epsilon^2$ arises from $Z'$ production, and the remaining terms originate from $\chi_2$ decays.
This indicates that increasing $m_{Z'}$ enhances the kinetic mixing, thereby causing both the relic density curve and the projected sensitivity regions to shift upward.
Furthermore, a larger mass ratio $m_{Z'}/m_{\chi_1}$ expands the projected sensitivity regions from lepton colliders, particularly at higher values of $m_{\chi_1}$.
And the pole feature in Fig.~\ref{bound6} appears around  $m_{\chi_1}\sim m_{Z'}/6\approx 15$GeV,
as the small value of $\Delta_\chi^5$ in Eq.(\ref{decay}) results in a longer observable lifetime for $\chi_2$.

\bibliographystyle{JHEP}
\bibliography{main}

\end{document}